\documentclass[12pt]{article}
\usepackage{latexsym}
\usepackage{graphicx}
\usepackage[usenames,dvipsnames]{color}

\def\be{\begin{equation}}
\def\ee{\end{equation}} 
\def\bea{\begin{eqnarray}}
\def\eea{\end{eqnarray}}

\def\e{\emph}

\def\nb{$N$-body problem }
\def\nbn{$N$-body problem}

\def\mb{$N$-body }

\def\case#1/#2{\textstyle\frac{#1}{#2}}

\def\e{\emph}

\def\n{Newtonian }

\begin{document}

\begin{center}

{\bf \Huge{Time's Arrow and Simultaneity:\\ A Critique of Rovelli's Views}}

\vspace{.3in}

{\bf \Large{Julian Barbour}}\footnote{julian.barbour@physics.ox.ac.uk.}

\end{center}

\abstract{In the joint paper ``Bridging the neuroscience and physics of time'' Rovelli, as the physicist coauthor of neuroscientist Dean Buonomano, makes statements that rely on theoretical frameworks employed when the laws of thermodynamics and general relativity were discovered. Their reconsideration in the light of subsequent insights suggests growth of entropy is not the origin of time's arrow and that a notion of universal simultaneity may exist within general relativity. This paper is a slightly extended form of my invited contribution to the forthcoming Frontiers in Psychology special issue ``Physical time within human time''.

\section{Introduction}

The great discoveries in physics---Copernican revolution, Newton's laws, thermodynamics, quantum mechanics, and relativity---were all made in a prevailing framework of concepts and knowledge. Thus, forces played no role in Copernicus's reformulation of Ptolemy's astronomy; it was first Kepler and then Newton who introduced them. For his part, Newton formulated his laws in the now rejected framework of absolute space and time. It was Carnot's study of steam engines that led to the discovery of thermodynamics in 1850, when virtually nothing was known about the universe at large. Finally, the spacetime form in which Einstein created general relativity obscured critical aspects of its dynamical structure that came to light decades later. The contexts in which these discoveries were made need to be taken into account in their modern interpretation and cast doubt on Rovelli's claims in \cite{grub}.

\section {Thermodynamics}

Of thermodynamics, Einstein said \cite{ein} ``It is the only physical theory of universal content which I am convinced that, within the framework of applicability of its basic concepts, will never be overthrown.'' Einstein did not spell out the basic concepts that ensure the applicability of thermodynamics. They are hard to find in textbooks, but history suggests what they are. Steam engines fail if their steam is not confined. Accordingly, for the atomistic explanation of the laws of thermodynamics the creators of statistical mechanics always assumed particles which collide with each other and bounce elastically off the walls of a box. Thermodynamics is the science of confined systems. In them growth of entropy (usually interpreted as a measure of disorder) from a special low entropy state is readily explained. What is impossible to explain by the known time-reversal symmetric laws of nature is the origin \e{in confined systems} of the special initial state.\footnote{It is important to distinguish systems that are \e{closed} in the sense of not being subject to external forces from ones that are confined, either physically by a box in a laboratory or theoretically by an artificial nondynamical potential wall. The \nbn, which I discuss in this paper, is closed in the traditional sense but not confined because critically, as I will emphasise, its scale variable can grow without bound. General relativity, which I also discuss, has spatially closed solutions, but their volume and Hubble radius grow without bound in eternally expanding solutions.}

In the theory of dynamical systems, those `in a box' correspond formally to ones that have a phase space of bounded Liouville measure; to my knowledge systems with phase spaces of unbounded measure have never been seriously studied in statistical mechanics. This is a remarkable lacuna since it is only in confined systems that the Poincar\'e recurrence theorem holds and is the main reason why a satisfactory explanation of time's arrow eluded Boltzmann \cite{bol} and so many since. We need to see if lifting the phase-space bound gives the \e{sine qua non} for solution to the problem of time's arrow: time-irreversible behaviour in generic solutions of time-reversal symmetric equations without invocation of a special condition.

I will show this is the case in the oldest dynamical theory, the \n \nb of gravitating point particles of masses $m_i, i=1,\dots N$. It has an unbounded phase space since its scale variable, the root-mean-square length
\be
\ell_\textrm{\scriptsize rms}={1\over M}\sqrt{\sum_{i<j}m_im_jr_{ij}^2},~~r_{ij}=|{\bf r}_i-{\bf r}_j|,~~M=\sum_i^Nm_i\label{rms},
\ee
is unbounded above. As shown in \cite{bkm,jp,fl} this leads to behaviour quite different from confined systems. If the energy is non-negative and (\ref{rms}) does not vanish, then as a function of Newton's absolute time $t$  the graph of $\ell_\textrm{\scriptsize rms}(t)$ is concave upwards and tends to infinity either side of a unique minimum or `Janus point'. Moreover, $D=\textrm d\ell_\textrm{\scriptsize rms}/\textrm dt$ is monotonic, tending from $-\infty$ to $\infty$ in both time directions; $D$ is therefore a Lyapunov variable, which rules out Poincar\'e recurrence: \mb statistics cannot be `box statistics'.

This is confirmed by the behaviour of the Newton potential energy
\be
V_\textrm{\scriptsize New}=-\sum_{i<j}{m_im_j\over r_{ij}}
\ee
made scale-invariant as the \e{shape potential} $V_\textrm{\scriptsize shape}$ through multiplication by  $\sqrt{\ell_\textrm{\scriptsize rms}}$:
\be
V_\textrm{\scriptsize shape}=\sqrt{\ell_\textrm{\scriptsize rms}}V_\textrm{\scriptsize New}.
\ee
With its sign reversed $V_\textrm{\scriptsize shape}$ may, as in \cite{bkm, jp}, be called the \e{shape complexity} since, independently of gravitational theory, it is a sensitive measure of the extent to which mass points in Euclidean space have a uniform or clustered distribution. Indeed $\ell_\textrm{\scriptsize mhl}^{-1}\propto-V_\textrm{\scriptsize New}$, where
\be
\ell_\textrm{\scriptsize mhl}^{-1}={1\over M^2}\sum_{i<j}{m_im_j\over r_{ij}}\label{mhl}
\ee
is the \e{mean harmonic length}, which means that
\be
C_\textrm{\scriptsize shape}={\ell_\textrm{\scriptsize rms}\over\ell_\textrm{\scriptsize mhl}}\label{comp}
\ee
has an absolute minimum when the particles are most uniformly distributed and increases rapidly when they cluster because that changes $\ell_\textrm{\scriptsize rms}$ relatively little but greatly reduces $\ell_\textrm{\scriptsize mhl}$. Note that $C_\textrm{\scriptsize shape}$ depends only on the \e{shape} of the particle distribution. In fact, the entire objective content in any  \mb solution resides entirely in the shapes it contains.\footnote{Imagined as `snapshots' showing the relative positions of the particles, the shapes could all be jumbled up in a random heap without any physical information being lost.} For the `shape-dynamic' representation of gravitational theory see \cite{fl}.

Provided $E\ge 0$ and zero-measure solutions to be discussed below are excluded, the qualitative behaviour of $C_\textrm{\scriptsize shape}$ shown in Fig.~1 for a numerical \mb solution of 1000 particles is universal: $C_\textrm{\scriptsize shape}$ grows with fluctuations between linearly rising bounds either side of the Janus point, where (as in the `artist's impression') the particle distribution is uniform and becomes clustered either side. The existence of the Janus point owes nothing to a special condition added to Newton's laws. It is present in every generic solution with $E\ge 0$.\footnote{Figure~1 (from \cite{bkm}) is for equal-mass particles with vanishing energy and angular momentum, $E={\bf L}=0$. These Machian conditions \cite{bb} match the dynamical structure of spatially closed general-relativistic solutions \cite{fl} and merely enhance the Janus-point behaviour always present when $E\ge 0$.} Further theories with unbounded phase spaces, other potentials, and Janus points in all of their solutions are listed in \cite{ad}.

\begin{figure}
\begin{center}
\includegraphics[width=0.8\textwidth]{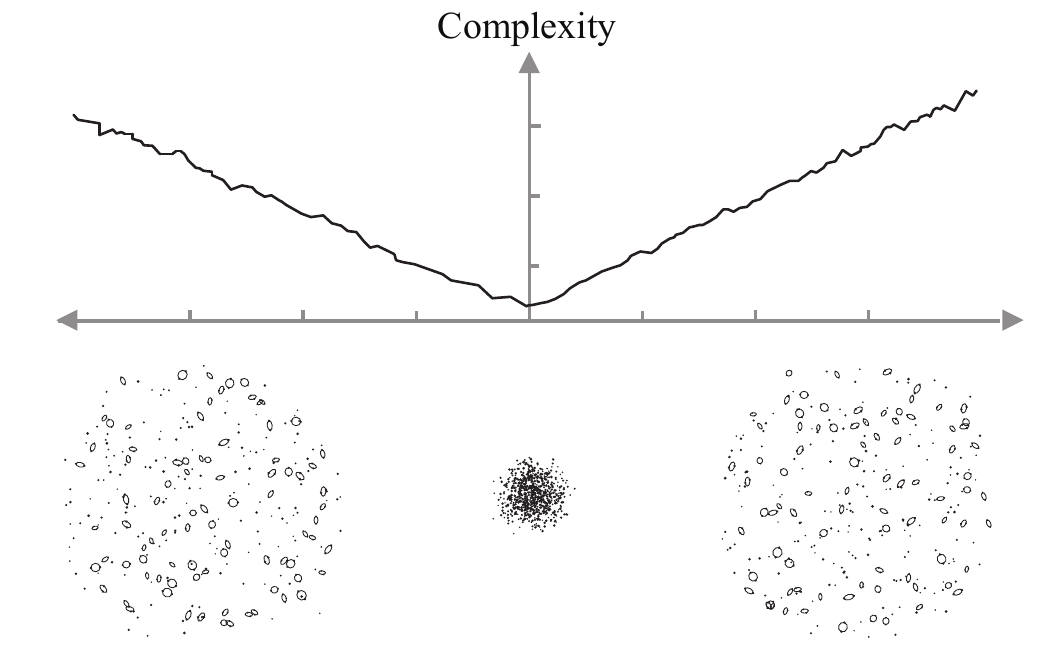}

Fig.~1 (from \cite{bkm}).
\end{center}
\end{figure}

The behaviour either side of the Janus point justifies its moniker: on the \mb timeline it lies between bidirectional arrows of time that the secular growth of $C_\textrm{\scriptsize shape}$ defines. Observers must be present on one or other side of the Janus point and even though Newton's laws are time-reversal symmetric will experience an arrow of time and an apparent birth of time and their universe in a uniform state in their past. The arrow does not arise from statistical behaviour of an ensemble of solutions and is secular, not a fluctuation. What is more the increased clustering that  $C_\textrm{\scriptsize shape}$ measures reflects the formation of more or less isolated systems that virialise and then generally decay. Their `birth-virialisation-death' arrows, the hallmark of self-confined thermodynamic systems like stars, all align with the master complexity arrow. It is noteworthy that subsystem decay often leaves two particles in elliptical orbits around their centre of mass. These `Kepler pairs' generally live forever and become ever better rods, clocks and compasses all in one, each of them perfectly synchronised with all the others \cite{bkm, jp}. Unlike growth of entropy, which is associated with increase of disorder, growth of complexity reflects increase of order. 

It is particularly interesting that the zero-measure \mb solutions mentioned above (and discussed in \cite{jp}, chaps. 16--18) model `big bangs' in general relativity: they begin with zero size as measured by (\ref{rms}). In them the behaviour of $C_\textrm{\scriptsize shape}$ matches what is seen on either side of the Janus point in Fig.~1. Finally, the solutions with negative energy either have a `Janus region' with bidirectional arrows of time either side of it or remain permanently bound and are virialised within a bounded region as is typical of statistical-mechanical systems with phase spaces of bounded measure. Moreover the relational Machian form of \n gravity \cite{bb} that models GR allows only solutions with $E={\bf L}=0$ and ensures there are always dynamical arrows of time. 

Thus, the complete set of solutions of the \nb divide into those that never leave a bounded region of their phase space and those that explore an unbounded region. The former exhibit typical thermodynamic behaviour while the latter always have master dynamical arrows of time and secondary thermodynamic behaviour in emergent subsystems that the master arrow entails. This is all a direct consequence of Newton's laws. Despite being time-reversal symmetric, they are fully capable of having solutions that are time-asymmetric on one or both sides of a distinguished point that has a dynamical as opposed to \e{ad hoc} origin and show that time-reversal symmetric dynamics is perfectly compatible with a dynamical (not statistical) arrow of time.

Let me compare these rigorous results in the \nb with some of the comments of Rovelli, who represents the physics side of [1] and says: ``For a theoretical physicist \dots~the distinction between past and future requires thermodynamics, hence is statistical only.'' Further, from the time-reversal invariance of ``all elementary mechanical laws'' it follows that ``the manifest time orientation of the world around us can only be a macroscopic phenomenon that is accounted for in terms of the distinction between micro and macrophysics, and caused by low entropy in the early universe.'' Despite Einstein's caveat, there is no discussion on Rovelli's part of the conditions of applicability of the basic concepts of thermodynamics. Along with virtually all physicists who have written about the arrow of time he retains the conceptual framework created by the discovery of thermodynamics through the study of steam engines and the associated restriction to dynamical systems with phase spaces of bounded Liouville measure. If, as is certainly the case for a model \mb universe, the phase space of the universe is unbounded, we must surely be prepared to reexamine dogmas from the 1850s.

\section{General Relativity and Simultaneity}

In the form in which Einstein formulated GR in 1915, there is no distinguished notion of simultaneity in spacetime, which can be foliated completely freely by three-dimensional hypersurfaces provided they are spacelike. However, decades later important insights were gained into the structure of GR when treated as a dynamical theory of evolving curved three-dimensional geometry. In 1958, Dirac \cite{dirac} cast GR into its simplest possible Hamiltonian form by eliminating all nonphysical degrees of freedom. He came to the remarkable conclusion that this could be done ``\e{only at the expense of givng up four-dimensional symmetry}'' (Dirac's emphasis) and continued ``I am inclined to believe from this that four-dimensional symmetry is not a fundamental feature of the physical world.'' In contrast to Einstein's great insight that ``each individual solution [in classical GR] \dots~exhibits four-dimensional symmetry'' Dirac argued this need not apply in a quantum theory because ``the individual solution has no quantum analogue''. Therefore ``Hamiltonian methods [needed for quantisation], if expressed in their simplest form, \e{force one to abandon the four-dimensional symmetry}''. 

In 1959 Dirac \cite{dir} introduced a condition which ensured this by introducing a dynamically distinguished notion of universal simultaneity (by maximal slicing), but only valid for universes of infinite spatial extent. In the same year Arnowitt, Deser, and Misner (ADM) \cite{adm} published the first of a series of papers that also cast GR into Hamiltonian form but, by not insisting on the simplest possible form, did so in a way that left open four-dimensional symmetry. It was the ADM form that Astekhar, Rovelli, and Smolin adopted in their attempt to create a discrete form of quantum gravity they called loop quantum gravity (LQG). However, despite some early success, the key problem in LQG---the quantum treatment of what is called the Hamiltonian constraint---has defied resolution for more than 30 years. The difficulty arises in large part because ADM retained four-dimensional symmetry. More evidence that it might be abandoned with advantage comes from important work of York \cite{york} in the early 1970s on the initial-value problem in GR. This showed that \e{three-dimensional} spatial symmetry is an integral part of GR and introduced a notion of universal simultaneity (defined by hypersurfaces of constant mean extrinsic curvature) that, while also casting the dynamics of GR into maximally simple form, differs from Dirac's choice in being applicable if the universe is spatially closed and allows it to evolve.\footnote{My comments about Dirac's and York's work need amplification and qualification and strictly hold only for `two-sided spacelike infinitesimal slabs' of globablly hyperbolic spacetimes. However, in both cases they characterise the dynamical structure of GR when cast in its maximally simple form. From the group-theoretical point of view, they establish a close parallel with the solutions of the \nb that are maximally simple through being completely free of the effects of Newtonian absolute structures \cite{sau}.}

I do not claim Dirac's and York's work definitively restores universal simultaneity but it does suggest need for a qualification of Rovelli's statement ``Objectively defined global simultaneity surfaces are not defined in general relativity.'' Moreover, since: 1) the currently observed accelerated expansion of the universe suggests that, like the \nbn, it too has an unbounded phase space; 2) Dirac's and York's maximally simple forms of GR have group-theoretical dynamical structures closely analogous to those of the \mb big-bang solutions; 3) there are analogues of the complexity both for vacuum GR \cite{jp, yam} and GR coupled to a scalar field \cite{isen} it seems likely that, \e{pace} Rovelli, the arrow of time in our universe has a dynamical and not statistical origin.

Finally, the complexity (\ref{comp}) has remarkable properties with radical possibilities I have not been able to address in this note and are discussed in \cite{sau}.

{\bf Acknowledgements.} My thanks to Pooya Farokhi and Anish Battacharya for helpful discussions.

\end{document}